\documentstyle[11pt]{article}

\def\be{\begin{equation}}
\def\ee{\end{equation}}
\def\bea{\begin{eqnarray}}
\def\eea{\end{eqnarray}}

\def\nn{\nonumber \\}
\def\e{{\rm e}}

\def\maketitle{\thispagestyle{empty}\setcounter{page}0\newpage
                \renewcommand{\thefootnote}{\arabic{footnote}}
                  \setcounter{footnote}0}
\renewcommand{\thanks}[1]{\renewcommand{\thefootnote}{\fnsymbol{footnote}}
               \footnote{#1}\renewcommand{\thefootnote}{\arabic{footnote}}}

\renewcommand{\title}[1]{\begin{center}\Large\bf #1\end{center}\rm\par\bigskip}
\renewcommand{\author}[1]{\begin{center}\Large #1\end{center}}
\newcommand{\address}[1]{\begin{center}\large #1\end{center}}

\def\babs{\hrule\par\begin{description}\item{Abstract: }\it} 
\def\eabs{\par\end{description}\hrule\par\medskip\rm}
\renewcommand{\date}[1]{\par\bigskip\par\sl\hfill #1\par\medskip\par\rm}

\begin{document}

\title{
Some issues related to conformal anomaly induced effective action
}
\author{ Sergio Zerbini\thanks{zerbini@science.unitn.it}}
\address{
Dipartimento di Fisica, Universit\`a di Trento \\ 
and Istituto Nazionale di Fisica Nucleare \\ 
Gruppo Collegato di Trento, Italia}

\begin{abstract}

Some issues related to quantum anomaly induced effects due to  matter are 
considered. Explicit examples corresponding 
to quantum creation of d4 dilatonic AdS Universe and of 
d2 dilatonic AdS Black Hole (BH) are discussed. 
Motivated by holographic RG, in a similar
approach, it is shown that, starting from a 5 dimensional AdS Universe,  
4-dimensional de Sitter or AdS world is generated on 
the boundary of such Universe as a result of quantum effects.
A 5-dimensional brane-world cosmological scenario is also considered, where
the brane tension is not longer a free parameter, but its role is taken 
by quantum effects induced by the 4-dimensional conformal anomaly associated 
with 
conformal coupled matter. As a result, consistent quantum creations of 
De Sitter or AdS curved branes are possible.

\end{abstract}

\maketitle

\section{Introduction}

In this paper, we will review  the role of quantum 
anomaly induced effective action (see \cite{1}) 
 in the dynamical realization of
dilatonic AdS backgrounds in various dimensions. 
First, the 4d dilatonic classical gravity with $N$ dilaton coupled 
quantum fermions is considered. Solving the effective 
equations,  it follows that quantum corrected dilatonic AdS 
Universe may occur, even in situation where such classical 
solution was absent. However, the probability of generation 
of AdS Universe is significally less than of quantum creation 
of de Sitter Universe. 

Making use of the same techniques, the  dilatonic gravity with dilaton
coupled massless quantum fermions in two dimensions is investigated. 
The complete 
anomaly induced effective action is found. The quantum 2d 
dilatonic AdS BH, which was not existing on classical level, 
is constructed as the solution of the effective equations, 
namely the creation of quantum 2d AdS BH is shown to occur.

 From another side, within the well kown  AdS/CFT correspondence,
 where AdS background 
is introduced from the very beginning,  we discuss the holographic 
RG action leading to warped 
compactification (RS-like Universe) in the region where 
both sides of AdS/CFT correspondence (i.e. bulk and boundary) 
are still relevant. Then, using the same anomaly induced 
effective action, previously constructed, we suggest 
the way it should appear in the dynamics of five-dimensional 
world, on equal footing with 5d gravity action. As a result,
the dynamical effective equations could be solved realizing 
5d AdS Universe with warp factor. On the boundary of such 
Universe, the de Sitter (inflationary) world occurs, namely it is 
actually induced by quantum effects, without introducing the four-dimensional 
cosmological constant on brane by hands.

\section{Quantum instability of 4d AdS dilaton universe}

 Let us begin with  the  4-dimensional dilaton gravity theory, descibed by
\be
\nonumber
S=\int d^4x \sqrt{-g} \left[ 
 -\frac{1}{16 \pi G}(R+ 6\Lambda) +\alpha
(\nabla_\mu \phi)(\nabla^\mu \phi) \right]\,,
\ee
where $\Lambda$ is the cosmological constant and $\alpha$ some suitable 
parameter. For constant dilaton $\phi$ and negative cosmological constant, 
the classical background solution corresponds to the 4d AdS space. 
Even for non-constant dilaton, there are solutions interpolating 
between asymptotically AdS and flat space with singular dilaton \cite{2}. 
Our primary 
interest will be  the issue of  the stability of the
(classical) 
AdS background in the above theory, under the quantum fluctuations of
the 
conformal matter. As  matter Lagrangian, we consider the one  corresponding
to $N$ massless dilaton coupled Dirac spinors
\be
\nn
L_M=\exp{(A \phi)}\sum_{i=1}^N \bar\psi_i\gamma^\mu\nabla_\mu\psi^i\ .
\ee
Here $A$ is some constant parameter.
The above strong matter-dilaton coupling is typical  for any matter- 
Brans-Dicke theory  in the Einstein frame.

The quantum effective action for dilaton coupled spinor has been found in 
ref. \cite{3} by integrating the conformal anomaly. This quantum effective 
action should be added to the classical one $S$ (there is only the 
dilaton-gravitational background under consideration).

Let us now define the space-time we are going to work with. 
We consider the 4d AdS background with the static metric 
\be
ds^2=\e^{-2\lambda \tilde{x}_3} \Big[dt^2-(dx_1)^2-(dx_2)^2 \Big]-
(d\tilde{x}_3)^2\,.
\label{3}
\ee
It has a  negative cosmological constant $\Lambda=-\lambda^2$. Making the 
coordinates transformations 
\be
y=\frac{\e^{\lambda \tilde{x}_3}}{\lambda} 
\ee
one can present (\ref{3}) in the conformally flat form  
\be
ds^2=a(y)^2 \eta_{\mu \nu}dx^{\mu}dx^{\nu}\,,
\label{33}
\ee
where
\be
a=\e^{-\lambda \tilde{x}_3}=\frac{1}{\lambda y}\,.
\ee

In order to investigate  the role of the quantum effects 
to the dilaton AdS universe, we shall consider the 
metric  (\ref{33})  with an arbitrary scale factor to be determined 
dynamically.
The anomaly induced effective action of ref. \cite{3} on such background 
may be written in the following form \cite{4}:

\be
\label{1ef}
W=V_3\int dy \left\{2b_1 \sigma_1 \sigma_1''''
 - 2(b + b_1)\left( \sigma_1'' - {\sigma_1'}^2 \right)^2\right\}\ .
\label{7}
\ee
Here, $V_3$ is the (infinite) volume of 3-dimensional flat space-time( time
is included), 
 $\sigma=\ln a(y)$, $\sigma_1=\sigma+\frac{A\phi}{3}$,
$'\equiv{d / dy}$, and
$b={3N \over 60(4\pi)^2}$, $b_1=-{11 N \over 360 (4\pi)^2}$.
It should be noted that (\ref{7}) may be regarded as the complete one-loop 
effective action.
The classical  gravitational action on the background (\ref{3}) with 
non-trivial dilaton reads
$$
S=V_3 \int dy \left\{ \frac{1}{\chi}\left[ 6(\sigma''+{\sigma'}^2) 
\e^{2\sigma}-6\Lambda \e^{4\sigma} \right]
+\alpha {\phi'}^2 \e^{2\sigma} \right\}\,.
\nn
$$
with $\chi=16 \pi G$. The quantum corrections of the conformal coupled matter 
field can be taken into account 
starting from the effective equations obtained from the effective action
$S+W$, thus implementing the standard semiclassical approach based on the 
Einstein Eqs. having as source the vacuum expectation value of the stress 
tensor of the quantum matter. These equations look similarly to the ones  
associated with the 
De Sitter quantum corrected universe \cite{4}, with the addition of the 
cosmological constant contribution and an opposite sign in the Einstein term:
One has \cite{5}
\bea
&&\tilde{C} \e^{({A \phi}/{3})}-\frac{12}{\chi}a''-\frac{24\Lambda a^3}{\chi}
 +2\alpha a\phi'^2=0 \nn 
&& \frac{A}{3}\tilde{C} a \e^{({A \phi}/{3})}-2\alpha (a^2 \phi')'=0\,,
\label{8}
\eea
where
\bea
\tilde{C}&=&
-\frac{4b}{\tilde{a}} \left[ \frac{\tilde{a}''''}{\tilde{a}}
-\frac{4\tilde{a}' \tilde{a}'''}{\tilde{a}^2} 
- \frac{3\tilde{a}''^2}{\tilde{a}^2}\right]\nn \\ 
& -&\frac{24}{\tilde{a}^4} \left[ (b-b_1) \tilde{a}'' \tilde{a}'^2 
 + b_1\frac{\tilde{a}'^4}{\tilde{a}} \right] \,, 
\label{88}
\eea
and
\be
\tilde{a}=a  \e^{({A \phi}/{3})}\,. 
\ee
First, it is easy to show that the  AdS Universe  exists at classical 
level.

Furthermore, in the absence of the dilaton and vanishing cosmological
constant,
only the first of (\ref{8}) survives. It may be solved via the special 
ansatz $a=c/y$ with the constraint $c^2=b_1\chi$. Since $b_1 <0$, one gets
an imaginary scale factor $a$, namely a quantum annihilation of the AdS 
universe, as it was shown in detail  in ref. \cite{6}. Hence, classical AdS 
Universe is not stable under the action of quantum fluctuations. Note that 
dilaton is constant there.

The general solution of the system of differential equations (\ref{8}) is 
very difficult to find. However, there  exist specific solutions \cite{5} 
\be
a(y)=\frac{1}{H y}\,,\,\,\,\,\,\, 
\phi'(y)=\frac{1}{H_1 y}\,,
\label{10}
\ee
where $H$ and $H_1$ are constants such that 

\be
\frac{12 \Lambda}{\chi H^2}= \frac{ \alpha}{H_1^2}
 -\frac{9 \alpha}{A H_1}-\frac{12}{\chi}\,. 
\label{11}
\ee
\be
\frac{81 \alpha}{2A b H^2}= -(A-3H_1)
\left(\frac{18H_1^2- 2\tilde{b}(A-3H_1)^2 }{H_1^2}\right)\,, 
\label{16}
\ee
with $\tilde{b}=1+\frac{b_1}{b}$. As a result, one may eliminate $H$ from the 
first equation and obtain a third  order complete algebraic equation for 
the quantity $H_1$, which in principle can be solved.
However, for $\Lambda=0$, there is the complete decoupling of the two 
equations and one easily arrives  at
\be
H_1=\frac{1}{24}\left[ -\frac{9 \alpha \chi}{A}\pm 
\sqrt{\frac{81 \alpha^2 \chi^2}{A^2}+ 48 \alpha \chi}\right]\,,
\label{17}
\ee 
and the other quantity $H$ may be obtained from equation (\ref{16}): 
\bea
\label{18}
\frac{81\alpha}{2b  H^2}&=&{2b \over 81\alpha}
\left(-18 + 81 \tilde b\right)A^2 
+ {24 \tilde b A^4 \over \alpha \chi}\nn \\
& \pm& {\alpha \chi \over 3\tilde b A}
\sqrt{\frac{81 \alpha^2 \chi^2}{A^2}+ 48 \alpha \chi}\ .
\eea
Since $\tilde b={7 \over 11}>0$ and $-18 + 81 \tilde b>0$, there is 
always a real solution for $H$ at least for positive $\alpha$ if we 
choose the sign $\pm$ in front of the square root properly. 

Hence, we demonstrated that in presence of non-constant dilaton 
the quantum AdS Universe solution in dilatonic gravity is less unstable.
At least, it may be realized while it did not exist on classical level!
The above mechanism may serve as
the one corresponding to quantum creation of primordial AdS Black 
Holes in early Universe. However,  for 4d AdS BH one should calculate
the extra piece of effective action which is non-local and very complicated.
The complete calculation of it is not known, presumbly it could be found 
only as expansion on theory parameters. That is the reason we prefer 
to present such analysis only in two dimensions.  

\section{ Quantum creation of 2d AdS black hole}

Here  we investigate the possibility for quantum creation
of 2d AdS BHs using methods developed in previous section.
Motivated by the 4-dimensional case, we may assume that the classical 
action  for the 2-dimensional dilaton gravity theory reads
\bea
S&=&\int d^2x \sqrt{-g(x)} \left[ -\frac{R+6\Lambda}{\chi}+\frac{1}{2}
(\nabla_\mu \phi)(\nabla^\mu \phi)\right]\nonumber  \\
&+&S_M \,,
\label{s2}
\eea
where $A$ is a constant parameter  and the matter 
Lagrangian is the one of two-dimensional Majorana spinors:
\bea
S_M&=&\int d^2x \sqrt{-g(x)}\exp{(A \phi)L_M}\nonumber \\
L_M&=&\sum_{i=1}^N \bar\psi_i\gamma^\mu\nabla_\mu\psi^i\ .
\eea
Let us neglect the classical matter contribution since we are interested 
only in the one-loop EA induced by the conformal anomaly of the quantum 
matter. 

In two dimensions, a general static   metric  may be written in  the form
\be
ds^2=V(r)dt^2-\frac{1}{W(r)} dr^2\,.
\label{st}
\ee
It is well know that introducing the new radial coordinate $r^*$, defined by
\be
r^*=\int \frac{dr}{\sqrt{V(r)W(r)}}\,,
\ee
one gets a conformally flat space-time, 
\be
g_{\mu \nu}=V(r(r^*))\eta_{\mu \nu}=\e^{2\sigma(r^*)}\eta_{\mu \nu}\,.
\ee
We also assume that the field  $\phi$ depends only on 
$r^*$.

Since the conformal anomaly for the dilaton coupled 
spinor is  (see \cite{3})
\be
T={c \over 2}\tilde R \ ,
\ee
where $c={N /( 12\pi)}$ and $\tilde R$ is calculated on the 
metric $\tilde g_{\mu\nu}=\e^{2A\phi}g_{\mu\nu}$, 
the anomaly induced EA  in the local, non-covariant form  reads 
\be
\label{eas}
W={c \over 2}\int d^2x \tilde\sigma \tilde\sigma'' \ .
\ee
Here $\tilde\sigma = \sigma + A\phi$ and $'=\frac{d}{dr^*}$. 

Note that this is, up to a
non--essential constant, an exact one-loop expression.  
The total one-loop effective action is $S+W$, i.e.
$$
S+W=
V_1\int dr^*  [\frac{k_G}{2} 
\left(\sigma''-6\Lambda\e^{2\sigma}\right)+\frac{1}{2} (\phi')^2] 
\nonumber 
$$
$$
+V_1\int dr^*  [\frac{c}{2}\left(\sigma + A\phi\right) 
\left(\sigma'' + A\phi''\right)  ] \,,
\nonumber
$$
where $k_G=\frac{1}{8 \pi G}$, and  $V_1$ the (infinite) 
temporal volume. 
Since 2d Einstein theory is trivial, the whole dynamics 
appears as a result of quantum effects.
The equations of motion given by the variations of $\phi$
and $\sigma$ are 
\bea
\label{eqm1}
0&=&-\left(1 - cA^2\right)\phi'' + cA \sigma'' \\
\label{eqm2}
0&=& c\left( \sigma'' + A \phi''\right) - 6k_G \Lambda 
\e^{2\sigma}\ .
\eea
 From (\ref{eqm1}) and (\ref{eqm2}), we obtain

\be
\label{eqm4}
E=-3k_G\Lambda \e^{2\sigma} 
+ {c \over 2\left(1 - cA^2 \right)}\left(\sigma'\right)^2\ .
\ee
where $E$ is a constant of integration. 

Thus, from   (\ref{eqm4}) one gets
\bea
\label{eqm7}
&&\e^{2\sigma}={(r-r_0)^2 \over b} - a \nn
&& b\equiv {6(1-cA^2)k_G \Lambda \over c}\ ,
\quad a\equiv {E \over 3 k_G \Lambda}\ .
\eea
Here $r_0$ is another constant of the integration. 
If we further redefine, 
\be
\label{eqm8}
{1 \over l^2} \equiv b \ ,\quad cM \equiv {2r_0 \over b}
\ ,\quad k \equiv -a + {r_0^2 \over b}\ ,
\ee
we obtain a generic 2d AdS black hole solution, 
\be
W(r)=V(r)=\e^{2\sigma}=k-cMr+\frac{r^2}{l^2}
\label{ads2}
\ee
where $M$ may be interpreted as the mass of the BH. We may take 
$k=\pm 1$, or $k=0$. In general, we have a simple positive root, 
interpreted as horizon radius. In the case $k=1$ one must 
have $cM >2$. It is easy to show that the above metric has 
a negative constant scalar curvature and for large $r$, 
$V(r) \simeq \frac{r^2}{l^2}$, namely one gets the AdS 
asymptotic behavior.  For the sake of simplicity, let us 
consider the case $k=0$. 
In this case the horizon radius and the Hawking temperature read
\be
\label{rHbH}
r_H=cMl^2\,\,\,, \beta_H=\frac{4 \pi}{cM}=\frac{4 \pi l^2}{r_H}\,. 
\ee

Thus, using anomaly induced effective action for dilaton coupled 
spinor we proved the possibility of quantum realization of 2d 
AdS BH which was not existing  at classical level. It is interesting 
that, unlike to 4d case where EA for AdS BH is not completely known, 
2d case is exactly solvable. Our solution may be interpreted as 
quantum creation of dilatonic AdS BH.

\section{Holographic Renormalization Group and Dynamical Gravity}

In the standard AdS/CFT correspondence, one can think 
about the simultaneous incorporation of string compactification 
with exponential warp factor (Randall-Sundrum compactificat 
 and the holographic map between
5d supergravity and 4d boundary (gauge) theory. Moreover, 
it could be extremely interesting to do it in such a way 
that dynamical gravity would appear on the boundary side. 
One possibility to realize such a mechanism is
presumbly related with holographic renormalization group (RG), see
\cite{7,8} for an introduction.

There was very interesting suggestion in this respect 
in ref.\cite{8} to consider low-energy effective 
action (EA) in the region where field theoretical 
quantities and analogous supergravity quantities 
could be considered on equal foot. In other 
words, this is the way to match two dual 
descriptions into the global picture of some, 
more universal RG flow. Immediate consequence of 
such point of view is the possible explanation of 
smallness of cosmological constant, the stability of 
flat spacetime along the RG flow and possible 
understanding of 4d gravity appearence in standard 
AdS/CFT set-up.

The explicit 
realization of ideas of ref.\cite{8} via the construction 
of the corresponding phenomenological model has been presented in 
\cite{5}. We summarize the results. 

In the calculation of complete low-energy EA in  
AdS/CFT correspondence,  one can divide it into a 
high energy and low energy pieces, separated by 
some given RG scale (fixed value of radial 
coordinate):
\be
\label{I1}
S=S_{\rm UV} + S_{\rm IR}\ .
\ee
Here $S_{\rm UV}$ is obtained from the original stringy 
action as a result of specific compactification. 
$S_{\rm IR}$ may be identified with the quantum effective 
action of (gauge and matter) low energy theory. 

Now let us consider 5d warped AdS metric: 
\be
\label{I2}
ds^2 =-dr^2+ a_1^2(r)\tilde g_{\alpha\beta}dx^\alpha dx^\beta 
\ee
where $r$ is radius of d5 AdS, or RS Universe,
i.e., $a_1 =a_1(r)$ is scale factor of d5 AdS and 
$a_1(r)$ usually depends exponentially on the 
radial coordinate. 
$\tilde g_{\alpha\beta}$ is 4d metric of boundary, 
time dependent FRW Universe. We assume that 
$\tilde g_{\alpha\beta}=a^2(\eta)\eta_{\alpha\beta}$ 
where $\eta$ is conformal time and 
$\eta_{\alpha\beta}$ is 4d Minkowski tensor. 
As $\tilde g_{\mu\nu}$ corresponds to conformally flat space,
it is defined by conformal time dependent scale factor $a$. 
Hence we have two scale factors.
 One can discuss now 
the structure of low-energy effective action. 
Truncation of $S_{\rm UV}$ gives basically the bosonic sector 
of 5d gauged supergravity and, for sake of  simplicity, 
we consider the situation with only a constant scalar (dilaton). 
As a consequence
\be
\label{I4}
S_{\rm UV}=\int d^5x\sqrt{-g_{(5)}}\left\{-{1 \over H}R_{(5)} 
 - {6\Lambda \over H}\right\}
\ee
where $V(\phi=\mbox{const})\equiv {6\Lambda \over H}$. 
We consider 5d AdS background with some 4d time-dependent 
conformally-flat boundary in this theory 
as vacuum state. The question 
is, can boundary quantum effects (instead of 4d cosmological constant)
stabilize such space?

The 4d quantum effective action of low-energy theory on 
the conformaly-flat space $g_{\mu\nu}=a^2(\eta)\eta_{\mu\nu}$ 
looks as (see, for example, \cite{6})
\be
\label{I5}
W=V_3\int d\eta \left\{2b_1 \sigma \sigma''''
 - 2 (b + b_1)\left(\sigma'' - {\sigma'}^2\right)^2\right\}
\ee
where $\sigma=\ln a(\eta)$, $V_3$ is space volume, 
$\sigma'={d \sigma \over d\eta}$, for ${\cal N}=4$ 
$SU(N)$ SYM theory $b={N^2 -1 \over 4(4\pi )^2}$, $b_1=-b$. 
In general, $b>0$, $b_1<0$ and $b\neq b_1$. 

One has to relate $W$ with $S_{IR}$. We 
consider d5 AdS background with the metric of the form:
\be
\label{I6}
ds_5^2 = -dr^2 + a_1^2(r)a^2(\eta)\eta_{\mu\nu}dx^\mu 
dx^\nu\ .
\ee

One knows that $S_{\rm IR}=W$ at $r=r_0$, cut-off 
scale. On the other side in AdS limit the 
description is completely from supergravity 
side. So, at $r\rightarrow r_A$, $S_{\rm IR}\rightarrow 0$. 
Then one can adopt the phenomenological 
approach where 
\be
\label{I7}
S_{\rm IR}=\int f\left(a_1(r)\right) W dr
\ee
so that $f\left(a_1(r)\right)$ satisfies above relations 
connecting $S_{\rm IR}$ and $W$ . A simple     
 choice for it
 is $  f\left(a_1(r)\right)=a_1(r)/a_1(r_0)$.    
Then, one can solve Eqs. of motion from 
$S_{\rm UV} + S_{\rm IR}$ on the background (\ref{I6}). 
As  result, the metric reads 
\be
\label{SS10}
ds^2 = -dr^2 + e^{\pm 2(r-r_0)/ l} ds_{\rm wall}^2\ .
\ee
where the metric on the wall of the brane is
\be
\label{dS2}
ds_{\rm wall}^2=dt^2 - \e^{2t}
\sum_{i=1}^3 \left(dx^i\right)^2\ ,
\ee
namely the metric of de Sitter space, 
which can be regarded as inflationary universe. It is interesting that 
Hubble parameter is depending from radial coordinate of 5d AdS Universe.
Therefore we have obtained the time dependent solution in the form of  
warped compactification, which is caused by the quantum correction 
coming from the boundary QFT. 
In the above treatment, we have assumed that the wall lies at 
$r=r_0$ since $f=1$ there. We need, however, to check 
the dynamics of the wall by solving junction equation coming 
from the surface counterterm, which should include $W$ 
 as a quantum correction. 

In the same way, when $b_1 <0$, one obtains as a solution, the Eq.
 (\ref{SS10}), but now  the metric of the wall of the brane is given by 
\be
\label{AdS}
ds_{\rm wall}^2={1 \over y^2}
\left( dt^2 - dy^2 - \left(dx^1\right)^2 
- \left(dx^2\right)^2 \right)\ .
\ee
The metric in (\ref{AdS}) is nothing but that of 4d AdS. Hence, one can get 
5d AdS Universe with warp scale factor a la Randall-Sundrum, where 
4d AdS world is generated on the wall. Again, as in section 2 the
probability of realization of 4d AdS is less than the one for de Sitter
Universe.  

In conclusion, we have exibit a  model where warped RS type scenario may 
be realized simultaneously with generation of inflationary 
Universe (or less stable AdS) on the wall. Dynamical 4d gravity is induced
from
background gravitational field on the boundary.
The source for such mechanism is quantum effects due to boundary 
QFT. 
It is not quite clear how one can estimate exactly these 
quantum effects.
That is the reason we adopted the phenomenological approach 
introducing some cut-off, interpolating, fifth coordinate 
dependent  function in such a way that near AdS the theory is 
described by 5d SG.  Far away of AdS,
at some fixed radius it is described by anomaly induced effective 
action of dual 4d QFT. There is, of course, some ambiguity in 
the choice of this function. However, that may be considered as 
kind of usual regularization dependence 
in frames of holographic RG.

\section{Quantum creation of a de Sitter (anti-de Sitter) 4 d universes}

Developing further the study of warped compactifications with curved 
boundary (inflationary brane) within AdS/CFT correspondence, the 
natural question is about the role of quantum bulk effects in 
such scenario.

This has been done in Ref. \cite{9}, where (at least, qualitatively) 
the role of bulk quantum effects to the scenario of 
refs.\cite{5,10,11} has been  considered and, consequentely,
 the  brane-world cosmology has been investigated. 
Here a summary of the results.

Let us start with a 5 dimensional bulk space  whose boundary is 
4-dimensional 
sphere S$_4$  or
4-dimensional hyperboloid H$_4$. 
The bulk metric  is given by 5 dimensional 
Euclidean Anti-de Sitter space $AdS_5$ 
\be
\label{AdS5i}
ds^2_{AdS_5}=dy^2 + \sinh^2 {y \over l}d\Omega^2_4\ .
\ee
Here $d\Omega^2_4$ is given by the metric of S$_4$ or H$_4$ 
with unit radius. One also assumes the boundary (brane) 
lies at $y=y_0$ 
and the bulk space is given by gluing two regions 
given by $0\leq y < y_0$ (see\cite{10} for more details.)

The starting point is the action $S$ which is the sum of 
the Einstein-Hilbert action $S_{EH}$, the Gibbons-Hawking 
surface term $S_{GH}$ \cite{12},  the surface counter term $S_1$ 
and the trace anomaly induced action $W$ 
\bea
\label{Stotal}
S&=&S_{EH} + S_{GH} + 2 S_1 + W \\
\label{SEHi}
S_{EH}&=&{1 \over 16\pi G}\int d^5 x \sqrt{g_5}\left(R_{5} 
+ {12 \over l^2}\right) \\
\label{GHi}
S_{GH}&=&{1 \over 8\pi G}\int d^4 x \sqrt{g_4}\nabla_\mu n^\mu \\
\label{S1}
S_1&=& -{3 \over 8\pi G}\int d^4 x \sqrt{g_4} 
\label{W}
\eea 
The factor $2$ in front of $S_1$ in (\ref{Stotal}) is coming from 
that we have two bulk regions which 
are connected with each other by the brane. 
In (\ref{GHi}), $n^\mu$ is 
the unit vector normal to the boundary. The expression for $W$ is omitted 
here and it is a complicated expression obtained form the 4 dimensional 
conformal anomaly, which depends on  
$G$  and $F$, the Gauss-Bonnet
invariant and the square of the Weyl tensor
\bea
\label{GF}
G&=&R^2 -4 R_{ij}R^{ij}
+ R_{ijkl}R^{ijkl} \nn
F&=&{1 \over 3}R^2 -2 R_{ij}R^{ij}
+ R_{ijkl}R^{ijkl} \ ,
\eea
and on the number of quantum fields by means of the coefficients $b$ and $b'$. 
For example, in the case of ${\cal N}=4$ $SU(N)$ super Yang-Mills theory 
$b=-b'={N^2 -1 \over 4(4\pi )^2}$. 
See  ref.\cite{11} for more details.

Motivated by (\ref{AdS5i}),
one assumes the following anzatz for 
the metric of 5 dimensional bulk space:
$$
\nonumber
ds^2=dy^2+e^{2A(y,\sigma)} l^2 (d\sigma^2+d\Omega^2_3)\,.
$$

The actions in (\ref{SEHi}), 
(\ref{GHi}), (\ref{S1}), and $W$, have the following forms:
\bea
S_{EH}&=& {l^4 V_3 \over 16\pi G}\int dy d\sigma \left\{\left( -8 
\partial_y^2 A - 20 (\partial_y A)^2\right)\e^{4A} \right. \nonumber \\
&& \left. +\left(-6\partial_\sigma^2 A - 6 (\partial_\sigma A)^2 
+ 6 \right)\e^{2A} + {12 \over l^2} \e^{4A}\right\} \nonumber \\
S_{GH}&=& {3l^4 V_3 \over 8\pi G}\int d\sigma \e^{4A} 
\partial_y A \\
\nonumber
S_1&=& - {3l^3 V_3 \over 8\pi G}\int d\sigma \e^{4A} \\
\nonumber
W&=& V_3 \int d\sigma \left[b'A\left(2\partial_\sigma^4 A
 - 8 \partial_\sigma^2 A \right) \right. \nn
&&\left. - 2(b + b')\left(1 - \partial_\sigma^2 A 
 - (\partial_\sigma A)^2 \right)^2 \right]\ .
\nonumber
\eea
Here $V_3=\int d\Omega_3$ is the volume or area of 
the unit 3 sphere.

However, 
 there is also the gravitational Casimir contribution due
to bulk quantum fields. It is possible to show, 
making use of zeta-function regularization techniques (see, for 
example \cite{13,14}),
 that for 
bulk scalar field, it has typically the following form $S_{\rm Csmr}$ 
\be
\nn
S_{\rm Csmr}={cV_3 \over {\cal R}^5}\int dy 
d\sigma \e^{-A} 
\ee
Note that role of (effective) radius of 4d constant curvature space
 is played by ${\cal R}\e^{A}$.
Here $c$ is some coefficient whose value and sign depend 
on the type of bulk field (scalar, spinor, vector, graviton, ...) 
and on parameters of bulk theory (mass, scalar-gravitational 
coupling constant, etc).  In the following discussion
it is more convenient to consider this coefficient to be some 
parameter of the theory.

Adding quantum bulk contribution
to the action $S$ in (\ref{Stotal}), the total action is
\be
\label{StotalB}
S_{\rm total}=S+S_{\rm Csmr}\,.
\ee
 ${\cal R}$ is 
the radius of S$_4$ or H$_4$. 

In the bulk, one obtains the effective equations of motion 
from $S_{EH} + S_{\rm Csmr}$ by the variation over $A$.
When scale factor depends 
on both coordinates:$y$,$\sigma$, 
one can find the solution of these Eqs. of motion as an expansion 
with respect to $\e^{-{y \over l}}$ by assuming that ${y \over l}$ 
is large, namely
\be
\nonumber
\e^A={\sinh {y \over l} \over \cosh \sigma}
- {32\pi Gcl^3 \over 15 {\cal R}^5}B^4(\sigma) 
\e^{-{4y \over l}} + {\cal O}\left(\e^{-{5y \over l}}\right)
\ee
for the perturbation from the solution where $B(\sigma)=\cosh \sigma$
for the S$_4$ brane  and $B(\sigma)=\sinh \sigma$ 
for  from H$_4$ brane.  

On the brane at the boundary, 
one gets the following equation: 
\bea
\label{eq2}
0&=&{48 l^4 \over 16\pi G}\left(\partial_y A - {1 \over l}
\right)\e^{4A}
+b'\left(4\partial_\sigma^4 A - 16 \partial_\sigma^2 A\right) \nn
&& - 4(b+b')\left(\partial_\sigma^4 A + 2 \partial_\sigma^2 A 
 - 6 (\partial_\sigma A)^2\partial_\sigma^2 A \right)\ .
\eea
We should note that the contributions from $S_{EH}$ and $S_{GH}$ are 
twice from the naive values since we have two bulk regions which 
are connected with each other by the brane. 
Substituting $e^A$  
into (\ref{eq2}), we find
$$
0\sim{1 \over \pi G}\left({1 \over {\cal R}}
\sqrt{k + {{\cal R}^2 \over l^2}}
+ {64\pi G l^7 c \over 3 {\cal R}^{10}}B^5(\sigma)
- {1 \over l}\right){\cal R}^4  
$$
$$
+ 8b'\ .
$$
where $k=1$ and $B(\sigma)=\cosh \sigma$ for $S_4$ brane and 
$k=-1$ and $B(\sigma)=\sinh \sigma $ 
for H$_4$ brane.
Here the radius ${\cal R}$ of S$_4$ or H$_4$ is related 
with $A(y_0)$, if we assume the brane lies at $y=y_0$, by
\be
\label{tldR}
\tilde R = l\e^{\tilde A(y_0)}\ .
\ee
In the above Eq,, only the leading terms 
with respect to $1/{\cal R}$ are kept in the ones coming from 
$S_{\rm Csmr}$ (the terms including $c$). 
When $c=0$, the previous result in \cite{10,11} is recovered. 
As a result,  
the Casimir force deforms the shape of S$_4$ or H$_4$ since 
${\cal R}$ becomes $\sigma$ dependent. The effect 
becomes larger for large $\sigma$. In case of 
S$_4$ brane, the effect becomes large if the distance from the 
equator becomes large. 
 Thus,  bulk quantum effects 
do not destroy the quantum creation of de Sitter (inflationary) 
or Anti-de Sitter brane-world Universe.

When $c=0$, the solution can exist when $b'<0$ for S$_4$ 
brane (in this case it is qualitatively similar to quite well-known
anomaly driven inflation of refs.\cite{15}) and $b'>0$ for H$_4$. 
For S$_4$ brane, if $b'<0$, the effect 
of Casimir force makes the radius smaller (larger) if $c>0$ ($c<0$).
For H$_4$ brane,  for small 
${\cal R}$ it behaves as 
\be
\label{H4slbr3}
0\sim {64 l^7 c \over 3 {\cal R}^{10}}\sinh^5\sigma
+ 8b'\ .
\ee
Then one would have solution even if $b'<0$. 

Furthermore, it is possible to show that  bulk quantum effects do not violate
(in some cases, even support) the quantum creation of de Sitter 
or Anti-de Sitter brane living in d5 AdS world.

For a recent review and a complete list of references, see  \cite{16} and 
for recent developments see
\cite{17,18}.

\section*{Acknowledgments}
{I would like to thank S. Nojiri and S.D. Odintsov for fruitful
collaboration}

\end{document}